\documentclass[fleqn]{article}
\usepackage{amsmath,amssymb,bm}
\begin{document}
\newcommand{\LCnabla}{\bar{\nabla}}
\newcommand{\FLCnabla}{\bar{\nabla}{}^F}
\newcommand{\Man}{\mathcal{M}}
\newcommand{\Lie}{\mathcal{L}}
\newcommand{\Intder}{\iota}
\newcommand{\SOdelta}{\delta_{SO(1,3)}}
\newcommand{\Mdual}[1]{\widetilde{#1}}
%
\title{Spinning particles in scalar-tensor gravity}
\author{D.A. Burton\thanks{Department of Physics, Lancaster
    University, UK}, R.W. Tucker$^*$ \& C.H. Wang\thanks{Department of
    Physics, National Central University, Taiwan}}
%
%
%
\maketitle
\begin{abstract}
We develop a new model of a spinning particle in Brans-Dicke
spacetime using a metric-compatible connection with
torsion. The particle's spin vector is shown to be Fermi-parallel
(by the Levi-Civita connection) along its worldline (an
autoparallel of the metric-compatible connection) when neglecting
spin-curvature coupling. 
\end{abstract}
%
%
%
%
\section{Introduction}
Scalar fields are replete in string-inspired low-energy effective
theories and occupy a prominent position in modern particle physics
and cosmology. The most widely accepted implementations of mass
generation and inflation employ scalar fields and it is not
unreasonable to suggest that they should play a role on
large scales. Indeed, a number of authors have suggested that
Einsteinian gravity (General Relativity), a purely metric-based theory, is incomplete and
should be augmented by a scalar component; one of the simplest
examples of such a theory was proposed by Brans and
Dicke~\cite{brans_dicke:1961}. Their theory was originally formulated as an action
principle whose degrees of freedom are the spacetime metric and the
Brans-Dicke scalar field $\varphi$. It was later
shown~\cite{dereli_tucker:1982} that Brans-Dicke theory could also be
obtained from an action whose independent variables are a
metric-compatible connection $\nabla$, an orthonormal frame field and
$\varphi$. Unlike the Levi-Civita connection $\LCnabla$ of
Einsteinian gravity, whose torsion vanishes identically, the field equations for
$\nabla$ yield a non-trivial torsion tensor in terms of $\varphi$.
Although the action principle
in~\cite{dereli_tucker:1982} differs from that introduced by Brans and
Dicke, the standard Brans-Dicke equations of motion are recovered when $\nabla$ is
expressed in terms of $\LCnabla$ and $\varphi$.

In Einsteinian and Brans-Dicke gravity,
electrically neutral spinless particles are \emph{postulated} to
follows autoparallels of $\LCnabla$ (geodesics). However,
the most natural connection in Brans-Dicke theory, from the
perspective of~\cite{dereli_tucker:1982}, is $\nabla$ and it
is perfectly reasonable to suggest~\cite{dereli_tucker:2002} that
electrically neutral spinless massive particles should follow the
autoparallels of $\nabla$ rather than $\LCnabla$. This hypothesis was
examined in~\cite{dereli_tucker:2002} where the autoparallels of
$\nabla$ and $\LCnabla$ were used to compare the perihelion shifts of Mercury
predicted by the two connections. The difference between the perihelion shifts predicted by
the new theory and the original Brans-Dicke theory are sufficiently
different to warrant further attention.

Analysis of inspiralling compact binaries using existing and future gravitational wave
detectors employs matched filtering techniques. Signal extraction relies
on accurate templates of the expected gravitational wave
emission and relativistic spin-orbit and spin-spin coupling play a
critical role (see~\cite{vecchio_et_al:2003} for a
recent discussion). Existing templates rely on post-Newtonian
treatments of Einsteinian gravity and may need modification if
the effects of torsion and spin are significant. Furthermore,
it is reasonable to suggest that the worldlines of spinning
particles could be governed by $\nabla$ rather than $\LCnabla$; the purpose
of this Letter is to investigate this possibility.  

Spinning particles in Einsteinian gravity have long
been a source of interest and consideration was recently given to
spinning particles in theories of gravity with
torsion~\cite{tucker:2004}. The general
approach followed in~\cite{tucker:2004} has two main ingredients :
(1) a set of relativistic balance laws and (2)
a set of constitutive relations (akin to equations of state in gas
dynamics).
The balance laws
are Noether identities arising from the diffeomorphism and gauge invariances of
an action; they only involve \emph{sources} (stress-energy-momentum tensor, spin tensor
and other currents). The constitutive relations are model-dependent equations for the
sources in terms of the true dynamical degrees of freedom and are
needed to reduce the balance laws to a closed system of field equations.

Many workers (see~\cite{tucker:2004} for a
review) have invested effort in deriving the equations of motion for a
spinning particle in General Relativity using the simplest source
models. In this Letter, motivated by~\cite{dereli_tucker:2002}, we
develop equations of motion for a spinning particle, based on a simple
source model, that lead to the autoparallel equation of $\nabla$ and the parallel
transport law with respect to $\nabla$. The
equations for a spinning particle presented here are a natural
alternative to the hypothesis that the Papapetrou-Dixon equations in
Einsteinian gravity are also valid in the context of Brans-Dicke gravity.
\section{Noether identities}
Balance laws are generated from diffeomorphism
and gauge invariances of an action. The approach adopted here
is a generalisation of~\cite{tucker:2004} and~\cite{benn:1982} to
accommodate the Brans-Dicke scalar field $\varphi$. For simplicity, we
will only consider electrically neutral matter and represent it using
a single\footnote{It is straightforward to
  include more fields and charged matter but this would draw
  attention away from the essential features of our argument.}
$p$-form $\Phi$.

Consider an effective action $S$
\begin{equation}
\label{action}
S[\Phi] = \int_\Man \Lambda
\end{equation}
for uncharged matter $\Phi$ in a
\emph{background} spacetime $\Man$ with metric $g$, metric-compatible
connection $\nabla$ and a \emph{background} Brans-Dicke scalar field $\varphi$.
The $4$-form $\Lambda$ is constructed tensorially from
$g,\nabla,\varphi,\Phi$ and, regardless of the detailed structure of
$\Lambda$, it follows
\begin{equation}
\Lie_X \Lambda \simeq \tau_a\wedge \Lie_X e^a + S_a{ }^b \wedge
\Lie_X \omega^a{ }_b + \rho\wedge\Lie_X\varphi +
\mathcal{E}\wedge\Lie_X\Phi
\end{equation}
where $\Lie_X$ is the Lie derivative with respect to any vector field
$X$ on $\Man$, $\simeq$
indicates equality up to an exact $4$-form, $\{e^a \}$ is a
$g$-orthonormal basis for $1$-forms and $\{\omega^a{ }_b\}$ are
the connection $1$-forms of $\nabla$ associated with $\{e^a \}$
($a,b,c = 0,1,2,3$). The
precise details of the sources (the stress $3$-forms $\tau_a$, spin
$3$-forms $S_a{ }^b$, 
$0$-form $\rho$ and $(4-p)$-form $\mathcal{E}$) depend on the details of
$\Lambda$; this will not concern us.

For a vector field $X$ with compact support it follows
\begin{equation}
\int_\Man \Lie_X \Lambda = \int_\Man \bigl( \tau_a\wedge \Lie_X e^a + S_a{ }^b \wedge
\Lie_X \omega^a{ }_b + \rho\wedge\Lie_X\varphi +
\mathcal{E}\wedge\Lie_X\Phi \bigr).
\end{equation}
Varying $S$ with respect to $\Phi$ yields
\begin{equation}
\delta S = \int_\Man \mathcal{E}\wedge\delta\Phi
\end{equation}
where $\delta\Phi$ is an arbitrary variation of $\Phi$ with compact
support. Thus, imposing the equations of motion $\mathcal{E}=0$ for $\Phi$ leads to
\begin{equation}
\label{Lie_Lambda}
\int_\Man \Lie_X \Lambda = \int_\Man \bigl( \tau_a\wedge \Lie_X e^a + S_a{ }^b \wedge
\Lie_X \omega^a{ }_b + \rho\wedge\Lie_X\varphi \bigr).
\end{equation}
Cartan's identity $\Lie_X\Lambda = \Intder_X d\Lambda + d
\Intder_X\Lambda$, where $d$ is the exterior derivative and $\Intder_X$ is the
interior product on forms with respect to $X$, yields
\begin{equation}
\int_\Man \Lie_X \Lambda = 0
\end{equation}
because $d\Lambda = 0$ and $X$ has compact support and using
(\ref{Lie_Lambda}) it follows
\begin{equation}
\label{diffeomorphism_invariance}
\int_\Man \bigl( \tau_a\wedge \Lie_X e^a + S_a{ }^b \wedge
\Lie_X \omega^a{ }_b + \rho\wedge\Lie_X\varphi \bigr) = 0.
\end{equation}
It can be shown~\cite{benn:1982}
\begin{align}
\label{Lie_e}
&\Lie_X e^a = D(\Intder_X e^a) + \Intder_X T^a - \SOdelta
e^a,\\
\label{Lie_omega}
&\Lie_X \omega^a{ }_b = \Intder_X R^a{ }_b - \SOdelta
\omega^a{ }_b
\end{align}
where $\SOdelta$ indicates an infinitesimal $SO(1,3)$ frame
transformation, $D$ is the exterior covariant derivative on the
orthonormal frame bundle, $T^a$ the torsion $2$-forms and $R^a{
}_b$ the curvature $2$-forms of $\nabla$ associated with $\{e^a\}$.
The action is invariant under $SO(1,3)$ frame transformations,
\begin{equation}
\label{SO_invariance}
0 = \int_\Man \SOdelta \Lambda = \int_\Man \bigl(\tau_a\wedge \SOdelta
e^a + S_a{ }^b \wedge \SOdelta \omega^a{ }_b \bigr),
\end{equation}
and it follows
\begin{equation}
\label{weak_noether_1}
\int_\Man \bigl(D\tau_c + \tau_a\wedge\Intder_{X_c}T^a +
S_a{ }^b \wedge \Intder_{X_c}R^a{ }_b +
\rho\,\Intder_{X_c}d\varphi\bigr) \mathcal{W}^c = 0
\end{equation}
using
(\ref{diffeomorphism_invariance}-\ref{SO_invariance})
and $X=\mathcal{W}^a X_a$ with
$\{X_a\}$ dual to $\{e^a\}$,
\begin{equation}
e^a(X_b) = \delta^a_b
\end{equation}
where $\delta^a_b$ is the Kronecker delta.
The action of $\SOdelta$ on $e^a$ and $\omega^a{ }_b$ is
\begin{align}
\label{SOdelta_e}
&\SOdelta e^a = -\mathcal{W}^a{ }_b e^b,\\
\label{SOdelta_omega}
&\SOdelta \omega^a{ }_b = D\mathcal{W}^a{ }_b
\end{align}
where $\mathcal{W}^a{ }_b$ is an element of the Lie algebra
$so(1,3)$ and has compact support on $\Man$. Using (\ref{SO_invariance}), (\ref{SOdelta_e}) and
(\ref{SOdelta_omega}) it follows
\begin{equation}
\label{weak_noether_2}
\int_\Man\biggl[DS_a{ }^b - \frac{1}{2}(\tau_a\wedge e^b -
\tau^b\wedge e_a)\biggr]\mathcal{W}^a{ }_b = 0
\end{equation}
and since (\ref{weak_noether_1}), (\ref{weak_noether_2}) hold for all
$\mathcal{W}^a$ and $\mathcal{W}^a{ }_b$ with compact support we
obtain the Noether identities
\begin{align}
\label{stress_identity}
& D\tau_c + \tau_a\wedge\Intder_{X_c}T^a +
S_a{ }^b \wedge \Intder_{X_c}R^a{ }_b +
\rho\,\Intder_{X_c}d\varphi = 0, \\
\label{spin_identity}
& DS_a{ }^b - \frac{1}{2}(\tau_a\wedge e^b -
\tau^b\wedge e_a) = 0
\end{align}
relating the sources $\tau_a, S_a{ }^b, \rho$.
\section{Equations of motion for a spinning particle}
Brans-Dicke theory can be obtained from a variational principle whose
independent variables are $\{e^a,\omega^a{
}_b,\varphi\}$~\cite{dereli_tucker:1982} and leads to the torsion $2$-forms
\begin{equation}
\label{torsion}
T^a = \frac{1}{2}T^a{ }_{bc}\,e^b\wedge e^c = e^a \wedge \frac{d\varphi}{\varphi}
\end{equation}
for $\varphi \neq 0$. Equations
(\ref{stress_identity}-\ref{torsion})
must be supplemented by further information
in order to obtain a closed system. The
following simple constitutive relations
\begin{align}
\label{stress_model}
&\tau^a = \varphi\,P^a \star e^0,\\
\label{spin_model}
&S_a{ }^b = \Sigma_a{ }^b \star e^0,\\
\label{rho_model}
&\rho = -P^0 \star 1
\end{align}
reduce to the model in~\cite{tucker:2004} when $\varphi$ is
constant,
where $\star$ is the Hodge map associated with $g$ and $\star 1$ is the
spacetime volume $4$-form. 

Equations describing a neutral spinning particle follow by taking
moments of (\ref{stress_identity}) and (\ref{spin_identity}) in
Fermi-normal coordinates on an open set $\mathcal{U}\subset\Man$
containing the image of the particle's worldline $\sigma$ with proper
time $t$ (the unit tangent to $\sigma$ is denoted $\dot{\sigma}$). The
orthonormal co-frame $\{e^a\}$ is defined on $\sigma$ such
that\footnote{$\Mdual{\dot{\sigma}}(Y) = g(\dot{\sigma},Y)$ for all
  vectors $Y$.}    
$e^0=-\Mdual{\dot{\sigma}}$ and $\{e^1,e^2,e^3\}$ are
Fermi-parallel (with respect to $\nabla$) along $\sigma$. Furthermore,
$\{e^a\}$ is
induced away from $\sigma$ by parallel transport along radial spacelike
autoparallels of $\nabla$ whose tangents are orthogonal to
$\dot{\sigma}$ (see~\cite{tucker:2004},~\cite{wang:2006} for details). Moments of
(\ref{stress_identity}) and (\ref{spin_identity}) are computed in
Fermi-normal coordinates $\{x^1,x^2,x^3\}$
on $\mathcal{U}$ and to leading order
\begin{align}
\label{dP0_dt}
&\frac{d \hat{P}^0}{d t} = -\bm{A}\cdot\bm{P},\\
\label{dPmu_dt}
&\frac{d \bm{P}_\mu}{d t} = -
  \frac{1}{\hat{\varphi}}\widehat{\frac{\partial\varphi}{\partial t}}\bm{P}_\mu -
\bm{A}_\mu\,\hat{P}^0 - \hat{T}^a{ }_{0\mu}\hat{P}_a -
\hat{P}^0\frac{1}{\hat{\varphi}}\widehat{\frac{\partial\varphi}{\partial x^\mu}}
+ \frac{1}{\hat{\varphi}}\hat{R}_{ab\mu 0}\hat{\Sigma}^{ab},\\
\label{dh_dt}
&\frac{d\bm{h}}{d t} = \bm{A}\times\bm{s} +
\frac{1}{2}\hat{\varphi}\bm{P},\\
\label{ds_dt}
&\frac{d\bm{s}}{d t} = \bm{h}\times\bm{A}
\end{align}
where
\begin{align}
&R^a{ }_b = \frac{1}{2}R^a{ }_{bcd} e^c\wedge e^d,\\
&\bm{A}_\mu = \hat{e}_\mu(\nabla_{\dot{\sigma}}\dot{\sigma}),
\end{align}
and hats indicate evaluation over the image of $\sigma$, i.e. at
$\{x^\mu = 0\}$, and
\begin{align}
&\bm{P}_\mu = \hat{P}_\mu,\\
&\bm{h}_\mu = \hat{\Sigma}^0{ }_\mu,\\
&\bm{s}_\mu = \frac{1}{2}\epsilon_{\mu\nu\omega}\hat{\Sigma}^{\nu\omega}
\end{align}
where $\epsilon_{\mu\nu\omega}$ is the Levi-Civita alternating
symbol with $\mu,\nu,\omega = 1,2,3$.
For given external fields $\varphi$ and $R^a{
}_{bcd}$,
equations
(\ref{dP0_dt}-\ref{ds_dt}) and (\ref{torsion})
are not sufficient to determine the
worldline and spin of the particle. Thus, the above system is supplemented
by the Tulczyjew-Dixon (subsidiary) conditions
\begin{equation}
\hat{P}_a\hat{\Sigma}^{ab} = 0
\end{equation}
i.e.
\begin{equation}
\label{TD_condition}
\bm{h}\hat{P}^0 = \bm{s}\times\bm{P}.
\end{equation}
Furthermore, using (\ref{torsion}) equation (\ref{dPmu_dt}) can be simplified to
\begin{equation}
\label{dPmu_dt_simplified}
\frac{d \bm{P}_\mu}{d t} = -
\bm{A}_\mu\,\hat{P}^0
+ \frac{1}{\hat{\varphi}}\hat{R}_{ab\mu 0}\hat{\Sigma}^{ab}.
\end{equation}
Equations (\ref{dP0_dt}), (\ref{dPmu_dt_simplified}), (\ref{dh_dt}),
(\ref{ds_dt}) and (\ref{TD_condition}) are a differential-algebraic
system for the worldline and spin of a particle in a Brans-Dicke
background spacetime.

Consider the weak-field limit of the above theory where the
spin-curvature coupling in (\ref{dPmu_dt_simplified}) is neglected,
\begin{equation}
\label{dPmu_dt_spin_coupling_neglected}
\frac{d \bm{P}_\mu}{d t} = - \bm{A}_\mu\,\hat{P}^0.
\end{equation}
Equations (\ref{dP0_dt}), (\ref{dPmu_dt_spin_coupling_neglected}), (\ref{dh_dt}),
(\ref{ds_dt}) and (\ref{TD_condition}) have the particular solutions
\begin{align}
\label{solution_start}
&\bm{P}=0,\\
&\bm{h}=0,\\
&\bm{A}=0,\\
&\bm{s}_\mu =\text{const.},\\
\label{solution_end}
&\hat{P}^0 = \text{const.}
\end{align}
and it immediately follows from (\ref{solution_start}-\ref{solution_end}) 
\begin{align}
\label{autoparallel}
&\nabla_{\dot{\sigma}} \dot{\sigma} = 0,\\
\label{parallel_spin}
&\nabla_{\dot{\sigma}} S = 0,\\
\label{orthogonal_spin}
&g(S,\dot{\sigma}) = 0
\end{align}
where $S = \bm{s}^\mu \partial/\partial x^\mu$ is the particle's spin
vector.
Clearly, for such solutions $\sigma$ is an autoparallel of $\nabla$ and
$S$ is parallel-transported with respect to $\nabla$ along
$\sigma$. 

The metric-compatible connection $\nabla$ and the
Levi-Civita connection $\LCnabla$ are related as
\begin{equation}
\Mdual{Z}(\nabla_X Y) = \Mdual{Z}(\LCnabla_X Y) +
\frac{1}{2}\bigl[\Mdual{X}(T(Z,Y)) + \Mdual{Y}(T(Z,X)) + \Mdual{Z}(T(X,Y))\bigr]
\end{equation}
where the $(2,1)$ torsion tensor $T$ and the torsion $2$-forms $T^a$,
equation (\ref{torsion}), satisfy
\begin{equation}
e^a(T(X_b,X_c)) = \Intder_{X_c}\Intder_{X_b}T^a = \delta^a_b
\frac{X_c\varphi}{\varphi} - \delta^a_c \frac{X_b\varphi}{\varphi}.
\end{equation}
It follows (\ref{autoparallel}-\ref{orthogonal_spin}) can be
written\footnote{$g(\Mdual{d\varphi},Y) = d\varphi(Y)$ for all vectors $Y$.}
\begin{align}
\label{LC_Brans-Dicke_gyro_start}
&\LCnabla_{\dot{\sigma}} \dot{\sigma} = - \frac{\Mdual{d\varphi}}{\varphi} -
 \frac{d\varphi(\dot{\sigma})}{\varphi}\dot{\sigma}  ,\\
&\FLCnabla_{\dot{\sigma}} S = 0,\\
\label{LC_Brans-Dicke_gyro_end}
&g(S,\dot{\sigma})= 0
\end{align}
where $\FLCnabla_{\dot{\sigma}}S$ is the Fermi-Walker
derivative of the $\dot{\sigma}$-orthogonal spin vector $S$ along
$\sigma$,
\begin{equation}
\FLCnabla_{\dot{\sigma}} S = \LCnabla_{\dot{\sigma}}S -
\Mdual{S}(\LCnabla_{\dot{\sigma}}\dot{\sigma})\dot{\sigma}.
\end{equation}
Equations
(\ref{LC_Brans-Dicke_gyro_start}-\ref{LC_Brans-Dicke_gyro_end}) can
also be written
\begin{align}
\label{LC_Brans-Dicke_gyro_start_components}
&A^a =
-\bigg(g^{ab} + \frac{dx^a}{dt}\frac{dx^b}{dt}\bigg)
\frac{1}{\varphi}\frac{\partial\varphi}{\partial x^b},\\
&\frac{d S^a}{dt} + \Gamma^a_{bc}S^b\frac{dx^c}{dt} = g_{ab}A^a S^b,\\
&g_{ab}S^a\frac{dx^b}{dt} = 0
\end{align}
for the particle's world-line $x^a(t)$ and spin $S^a$ where
\begin{equation}
\label{LC_Brans-Dicke_gyro_end_components}
A^a = \frac{d^2 x^a}{dt^2} + \Gamma^a_{bc}\frac{dx^b}{dt}\frac{dx^c}{dt}
\end{equation}
and $\Gamma^a_{bc}$ are the Christoffel symbols of the
Levi-Civita connection induced by the metric $g_{ab}$.

Equations
(\ref{LC_Brans-Dicke_gyro_start_components}-\ref{LC_Brans-Dicke_gyro_end_components}) are a natural
generalisation to Brans-Dicke theory of the conventional model of a freely-falling ideal
gyroscope in Einsteinian gravity,
\begin{align}
&\LCnabla_{\dot{\sigma}} \dot{\sigma} = 0,\\
&\LCnabla_{\dot{\sigma}} S = 0,\\
&g(S,\dot{\sigma})= 0
\end{align}
i.e.
\begin{align}
\label{Einsteinian_start_components}
&A^a = 0,\\
&\frac{d S^a}{dt} + \Gamma^a_{bc}S^b\frac{dx^c}{dt} = 0,\\
\label{Einsteinian_end_components}
&g_{ab}S^a\frac{dx^b}{dt} = 0.
\end{align}
\section{Conclusion}
We have developed a simple model of a spinning particle on a
Brans-Dicke background. Brans-Dicke theory is naturally formulated in
terms of a metric-compatible connection $\nabla$ with torsion and the present
model, in the weak-field regime, exhibits solutions where the
particle's worldline $\sigma$ is an autoparallel of $\nabla$ and its
spin vector $S$ is parallel (with respect to $\nabla$) along
$\sigma$. In terms of the Levi-Civita connection $\LCnabla$, the worldline
$\sigma$ has non-zero acceleration and $S$ is Fermi-parallel along $\sigma$.

Detailed knowledge of the collapse of compact spinning
binaries is important for attempts to detect gravitational
waves. Furthermore, it is invisaged that gravitational radiation will be used
as a tool to study astrophysical objects and, given the plethora of
scalar fields in low-energy string-inspired field theories, an effective
Brans-Dicke scalar may play a significant role. Any predictive theory,
in the present context, of the behaviour of a spinning particle on a
Brans-Dicke background relies on a source model that leads to a
sensible Newtonian limit for the theory. The above (with $\varphi$ constant)
has the same Newtonian limit as the Papapetrou-Dixon equations in
Einsteinian gravity. 

The approach discussed here is a simple
generalisation of the source model discussed in~\cite{tucker:2004} and
further work is necessary to elucidate the differences between such
models when spin-curvature coupling is significant.

In the weak field limit, the particle behaves like an
accelerating ideal gyroscope in Einsteinian gravity.
It may be shown~\cite{wang:2006} that the precession
rates (geodetic and Lense-Thirring) in Kerr spacetime given by
(\ref{Einsteinian_start_components}-\ref{Einsteinian_end_components})
and the corresponding rates in Kerr-Brans-Dicke spacetime given by
(\ref{LC_Brans-Dicke_gyro_start_components}-\ref{LC_Brans-Dicke_gyro_end_components})
are indistinguishable to leading order.
This has implications for any attempt to use 
Gravity Probe B~\cite{buchman:2000} to test the novel theory.
\section{Acknowledgements}
CHW thanks Prof Hoi-Lai Yu for his hospitality during a stay at Institute of
Physics, Academia Sinica, Taiwan.

\end{document}